\documentclass[nohyper,12pt,letterpaper]{JHEP3}
\pdfoutput=1
\usepackage{graphicx,epsfig,youngtab}

\author{Robert de Mello Koch$^{1,2}$, Badr Awad Elseid Mohammed$^{1}$, Jeff Murugan$^{3,4}$ and Andrea Prinsloo$^{3,4}$\\
$^{1}$National Institute for Theoretical Physics,\\
Department of Physics and Centre for Theoretical Physics,\\ 
University of the Witwatersrand,\\ 
Wits, 2050, South Africa\\
\qquad\\
$^{2}$Stellenbosch Institute for Advanced Studies,\\
Stellenbosch, South Africa\\
\qquad\\
$^{3}$Astrophysics, Cosmology \& Gravity Center and\\
Department of Mathematics and Applied Mathematics,\\
University of Cape Town,\\
Private Bag, Rondebosch, 7700, South Africa\\
\qquad\\
$^{4}$National Institute for Theoretical Physics,\\
Private Bag X1,\\
Matieland, 7602, South Africa.\\
\qquad\\
E-mail: \email{robert@neo.phys.wits.ac.za, bmohamme@ictp.it, jeff@nassp.uct.ac.za, andy.prinsloo@uct.ac.za}}

\abstract{In this article we consider gauge theories with a $U(N)\times U(N)$ gauge group. We provide, for the first time, a 
complete set of operators built from scalar fields that are in the bi fundamental of the two groups. Our operators
diagonalize the two point function of the free field theory at all orders in $1/N$. We then use this basis to investigate 
non-planar anomalous dimensions in the ABJM theory. We show that the dilatation operator reduces to a set of decoupled harmonic 
oscillators, signaling integrability in a large $N$ but non-planar limit of the theory.}

\preprint{ACGC-120222,\,\, WITS-CTP-088}

\title{Beyond the Planar Limit in ABJM}

\keywords{D-branes, AdS/CFT correspondence, Giant gravitons, Integrability}

\newcommand{\bea}{\begin{eqnarray}}
\newcommand{\eea}{\end{eqnarray}}

\begin{document}

\section{Discussion and Conclusions}

The AdS/CFT correspondence\cite{Maldacena:1997re} provides a definition of quantum gravity on negatively curved spaces.
In this article we are interested in exploring ${\cal N}=6$ $U(N)\times U(N)$ Chern-Simons theory (the ABJM model\cite{ABJM,BL}) 
which is dual to the type IIA superstring on AdS$_4\times\mathbb{CP}^3$. The dictionary between the two corresponding theories 
is organized by identifying the conformal dimensions of the operator in the field theory with the energy of the corresponding
state in the dual quantum gravity. This idea leads to the identification of operators with dimension $\sim 1$ with gravitons,
of certain operators with dimension $\sim \sqrt{N}$ with string states, of certain operators with dimension $\sim N$ with $D$-brane 
states and of certain operators with dimension $\sim N^{2}$ with new geometries. Clearly, operators with dimension $\sim N$ or
$\sim N^2$ have the potential to provide deep and important lessons. The study of these operators is however highly non-trivial
because the large $N$ limit of their correlation functions is not captured by summing the planar diagrams\cite{Balasubramanian:2001nh}. 
One is concerned here with a large $N$ but non-planar limit.

In the last few years it has become possible to study large $N$ but non-planar limits of gauge theories with gauge group $U(N)$.
The new idea allowing progress is to exploit group representation theory to construct a basis of operators that diagonalize the
free field two point function exactly. By now there are a number of interesting results. The first paper\cite{Corley:2001zk} to 
employ these ideas used Schur polynomials to provide a basis for the half-BPS sector of ${\cal N}=4$ Super Yang-Mills Theory.
These Schur polynomials were generalized to provide a basis for multi matrix operators in \cite{Balasubramanian:2004nb,de Mello Koch:2007uu}.
The basis constructed in \cite{Kimura:2007wy,Kimura:2009wy} is built using elements of the Brauer algebra. 
A basis that has sharp quantum numbers for the global symmetries of the theory was given in \cite{Brown:2007xh,Brown:2008rr}.
Finally, for a rather general approach which correctly counts and constructs the weak coupling BPS operators see\cite{jurgis}.
The results obtained in \cite{jurgis} can be translated into any of the bases we have considered.

In this article we will argue that, for theories with a $U(N)\times U(N)$ gauge group, a basis of local operators that again diagonalizes  
the free field theory to all orders in $1/N$, can again be constructed. By restricting to what is a single matrix sector of the theory,
\cite{dey} were able to compute the correlation function of some operators in a large $N$ but non-planar limit. This work showed that by
recycling some ideas that worked for $U(N)$ gauge group, there are questions that can be answered for the theory with a $U(N)\times U(N)$ 
gauge group. Here we are after a genuine extension of the existing methods. For a $U(N)$ gauge group, two point functions of operators built
from $n$ Higgs fields reduce to evaluating a certain trace over $V^{\otimes n}$,
where $V$ is the carrier space of the fundamental representation of $U(N)$. In contrast to this, when one considers a theory with
$U(N)\times U(N)$ gauge group, two point functions of operators built from $n$ Higgs fields\footnote{Our Higgs fields are of two types:
the first type are in the $(N,\bar{N})$ of the gauge group; the second are in the $(\bar{N},N)$ of the gauge group. Hence any sensible
gauge invariant operator is built from traces of an alternating product of the two types.} reduce to evaluating a product of traces over 
$V^{\otimes n}$, one for each gauge group. Our main idea (see section \ref{generalcase}) is to construct our operators in such a way 
that one of these traces simply produces a delta function which identifies the way Wick contractions are performed on the two types of 
fields. This idea can easily be generalized to study the general problem of a gauge theory whose gauge group has an arbitrary number of
factors, as explained in the Appendix.

Given a complete basis, many interesting questions can be asked. One question which has proved rather fruitful, is
to compute anomalous dimensions in this large $N$ but non-planar limit\cite{Koch:2010gp,VinceKate,bhw,Koch:2011jk,dgm,deMelloKoch:2011ci}.
It is now beyond question that integrability has proven to be a powerful tool in analyzing ${\cal N} = 4$ super Yang-Mills theory 
in the planar limit\cite{Minahan:2002ve,Beisert:2010jr}. The corresponding results achieved for the large $N$ but non-planar 
limit\cite{Koch:2010gp,VinceKate,bhw,Koch:2011jk,dgm,deMelloKoch:2011ci} are encouraging and, with some optimism, one hopes that 
integrability will play an equally important role in this case too. Indeed, preliminary results suggest that the dilatation operator
reduces to a set of decoupled oscillators. Motivated by this, we have studied anomalous dimensions for a class of operators in our basis.
The operators we study belong to the $SU(2)\times SU(2)$ subsector that is not mixed with other subsectors by the dilatation operator. 
For this sector the dilatation operator reduces to two decoupled $SU(2)$ chains on the even and odd sites\cite{Minahan:2002ve}.
We will only excite one of these chains so that we are ultimately left with a single $SU(2)$ chain, which is exactly what one has
from the $SU(2)$ sector of ${\cal N}=4$ super Yang-Mills theory.
We thus expect that the spectral problem considered here will be very similar to that obtained from ${\cal N}=4$ super Yang-Mills theory,
a fact that we explicitely demonstrate.
The methods of \cite{dgm} are powerful enough to handle this situation. In this case we again find that the dilatation operator reduces
to a set of decoupled oscillators!

There are a number of interesting directions in which this work can be extended. We will describe at least a few of these possibilities.
The operators whose anomalous dimensions we have studied are AdS/CFT dual to giant gravitons\cite{McGreevy:2000cw}. Giant gravitons have been
constructed in the AdS$_4\times\mathbb{CP}^3$ geometry in \cite{Jeff}. For related relevant studies see \cite{AdS4likegiants}. 
It would be interesting to explore this connection further and to see how much of the physics of the giant graviton can be recovered from
our operators. See \cite{jurgissanjaye} for very recent progress on developing the map between quantum states and brane geometries.
The current setting may be an interesting laboratory for these questions. Indeed, in the AdS$_5\times$S$^5$ setting a restricted
Schur polynomial with long rows corresponds to a collection of D3-dipole branes that each wrap an S$^3\subset$ AdS$_5$; a restricted
Schur polynomial with long columns corresponds to D3-dipole branes that each wrap an S$^3\subset$ S$^5$. For the problem we study here
a restricted Schur polynomial with long rows corresponds to D2-dipole branes that each wrap an S$^3\subset$ AdS$_4$ \cite{D2-dipoles}; 
a restricted
Schur polynomial with long columns corresponds to D4-dipole branes that each wrap a four dimensional manifold in $\mathbb{CP}^3$.

Secondly, when explicitly constructing the action of the dilatation operator, we have restricted ourselves to a class of operators dual 
to systems of two giant gravitons. The extension to systems of $p>2$ giant gravitons proved very fruitful\cite{dgm} in the AdS$_5\times$S$^5$
set up. It would be interesting to consider this generalization in the present setting. 

Finally, the action of the dilatation operator in the scaling limit that we consider in section \ref{scaling} reduces it 
to the action of the dilatation operator discovered in the $SU(2)$ sector of ${\cal N}=4$ super Yang-Mills theory. Can one find
other scaling for which the spectrum of anomalous dimensions can still be determined? Can one directly solve the resulting
recursion relations without taking a scaling limit?

\section{Some Notation}

We will denote the fields of the theory that we use as $A_1$, $A_2$, $B_1^\dagger$ and $B_2^\dagger$.
The number of $A_i$s is $n_i$; the number of $B_i^\dagger$s is $m_i$. Set $n=n_1+n_2=m_1+m_2$.
$R$ is an irrep of $S_n$, i.e. $R\vdash n$. Introduce the notation
$$
\phi_{11}{}^a_b =A_1{}^a_\alpha B_1^\dagger{}^\alpha_b \, ,
\qquad
\phi_{12}{}^a_b =A_1{}^a_\alpha B_2^\dagger{}^\alpha_b \, ,
$$
$$
\phi_{21}{}^a_b =A_2{}^a_\alpha B_1^\dagger{}^\alpha_b \, ,
\qquad
\phi_{22}{}^a_b =A_2{}^a_\alpha B_2^\dagger{}^\alpha_b\, .
$$
The number of $\phi_{ij}$s is $n_{ij}$.
$r_{ij}\vdash n_{ij}$ is an irrep of $S_{n_{ij}}$.
The collection, $(r_{11},r_{12},r_{21},r_{22}) \equiv \{ r\}$ is an irrep of 
$S_{n_{11}}\times S_{n_{12}} \times S_{n_{21}} \times S_{n_{22}} \subset S_{n}$.
We will be exploiting the fact that general multi trace operators can be realized as 
a single trace over the larger space $V^{\otimes n}$. In $V^{\otimes n}$ 
permutations have matrix elements
$$
  \langle i_1,i_2,\cdots,i_n|\tau |j_1,j_2,\cdots j_n\rangle = \delta^{i_1}_{i_{\tau(1)}}
   \delta^{i_2}_{i_{\tau(2)}}\cdots \delta^{i_n}_{i_{\tau(n)}}\, .
$$

\section{A Complete Set of Operators}
\label{completeops}

Consider the most general gauge invariant operator built using an arbitrary number of
$A_1$s, $A_2$s, $B_1^\dagger$s and $B_2^\dagger$s. Due to the index structure of the 
fields, any single trace gauge invariant operator is given by an alternating sequence of
pairs of $A_i$s and $B^\dagger_i$s. The possible pairs are the $(\phi_{ij}){}^a_b$ defined above.
Any single trace gauge invariant operator is given by a unique (up to cyclic permutations)
product of the $(\phi_{ij})^a_b$. The most general gauge invariant operator is given by a product
of an arbitrary number of these single trace operators. In this section we will provide a
new basis for these operators.

The basis we have constructed is given by a restricted Schur polynomial in the $\phi_{ij}$
\begin{center}
  \fbox{
    \begin{minipage}[c]{15cm}
        {\vskip 0.05cm}
        \begin{eqnarray}
           \label{basicop}
           O_{R,\{ r\}}={1\over n_{11}! n_{22}!n_{12}!n_{21}!}
           \sum_{\sigma\in S_{n}} {\rm Tr}_{\{ r\}}\left(\Gamma_R (\sigma)\right)
           {\rm Tr} (\sigma \phi_{11}^{\otimes n_{11}}\phi_{12}^{\otimes n_{12}}
           \phi_{21}^{\otimes n_{21}}\phi_{22}^{\otimes n_{22}})\, .
        \end{eqnarray}
        {\vskip 0.05cm}
    \end{minipage}
  }
\end{center}
The irrep $R$ will in general be a reducible representation of the $S_{n_{11}}\times S_{n_{12}}\times S_{n_{21}}\times S_{n_{22}}$
subgroup of $S_{n}$. One of the $S_{n_{11}}\times S_{n_{12}}\times S_{n_{21}}\times S_{n_{22}}$ irreps that $R$ subduces is $\{ r\}$. 
In the above formula, ${\rm Tr}_{\{r\}}$ is an instruction to trace only over the $\{ r\}$ subspace of the carrier space of $R$. A 
very convenient way to implement this trace is as
$$
{\rm Tr}_{\{ r\}}\left(\Gamma_R (\sigma)\right)=
{\rm Tr}\left(P_{R, \{ r\}}\Gamma_R (\sigma)\right)
\equiv \chi_{R,\{ r\}}(\sigma )
$$
where $P_{R, \{r\}}$ is a projector which projects from the carrier space of $R$ to the $\{r\}$ subspace. To prove that
these operators form a basis, we simply need to show that they are complete. We will demonstrate completeness by
showing that the most general gauge invariant operator built using an arbitrary number of $A_1$s, $A_2$s, 
$B_1^\dagger$s and $B_2^\dagger$s can be written as a linear combination of the operators (\ref{basicop}).
Further, we can argue that all the operators in this set are linearly independent. This follows immediately
from the fact that the number of restricted Schur polynomials is equal to the number of gauge invariant operators
(which are linearly independent). This counting agreement was proved in \cite{Collins:2008gc} at both finite and infinite $N$.

Now for the demonstration: The most general gauge invariant operator that we are considering can be written as
$$
o(\tau)={\rm Tr}(\tau \phi_{11}^{\otimes n_{11}}\phi_{12}^{\otimes n_{12}}\phi_{21}^{\otimes n_{21}}\phi_{22}^{\otimes n_{22}})
$$
for a suitable choice of the permutation $\tau\in S_{n}$. The completeness of this
basis now follows from the identity (which is derived in \cite{Bhattacharyya:2008xy})
$$
o(\tau)=\sum_{R,\{r\}}
{d_R n_{11}!n_{12}!n_{21}!n_{22}!\over d_{r_{11}}d_{r_{12}}d_{r_{21}}d_{r_{22}}\, n!}
\chi_{R,\{ r\} }(\tau) O_{R,\{ r\}}
$$
where the sum over $R$ runs over all irreps of $S_{n}$ and $\{ r\}$ ranges over all irreps of 
$S_{n_{11}} \times S_{n_{12}} \times S_{n_{21}} \times S_{n_{22}}$. This completes the demonstration.

The operators given in (3.1) do not have simple two point functions and they are not orthogonal. For this reason we 
will need to generalize (3.1) (see formula (4.1) below). The main insight gained from this section is the fact that
the number of gauge invariant operators is equal to the number of distinct restricted Schur labels $R,\{ r\}$. In what follows,
we are able to construct a set of operators that have orthogonal two point functions and are labeled by $R,\{ r\}$. Given the
lesson of this section, we know they are complete.

\section{Two Point Functions}
\label{twopoints}

In this section we will study the two point correlation functions of the operators (\ref{basicop}).
These correlators provide an interesting generalization of the correlators considered for operators
built from complex Higgs fields transforming in the adjoint of a $U(N)$ gauge theory. For correlators
built from $n$ Higgs fields transforming in the adjoint of a $U(N)$ gauge group one is able to reduce the
computation of the correlator to the computation of a trace over the space $V^{\otimes n}$ where $V$
is the carrier space of the fundamental representation of $U(N)$. In the present case, because we consider
a theory with $U(N)\times U(N)$ gauge group, the computation of the correlator reduces to a product of
two traces (one for each $U(N)$ factor in the gauge group) each of which run over $V^{\otimes n}$. We
will explain how to explicitly compute this trace. The first result we obtain
in this section is a general formula for the two point correlation function. We then consider the explicit
evaluation of this general result in two special cases: when $n_{12}=n_{21}=0$ and when $n_2=0$. These 
special cases are simpler than the general result, and the case $n_2=0$ represents a class of operators that
are closed under the action of the two loop dilatation operator. We will study the anomalous dimensions of these
operators in a later section. The final result of this section is a general formula for two point functions.
In an Appendix we explain how this result generalizes to gauge groups with more factors.

As has become standard in computations of this type, we ignore spacetime dependence; it is uniquely determined
in the final result by conformal invariance. Consequently, the two point functions of the Higgs fields that we 
use are
$$
\langle A_i{}^a_\alpha A_j^\dagger{}^\beta_b\rangle
=\delta_{ij}\delta^a_b\delta^\beta_\alpha =
\langle B_i{}^a_\alpha B_j^\dagger{}^\beta_b\rangle
$$
In terms of these Higgs fields we can write our operator as
$$
O_{R,\{r\}}={1\over n_{11}! n_{22}!n_{12}!n_{21}!}
\sum_{\sigma\in S_{n}}
{\rm Tr}_{\{r\}}\left(\Gamma_R (\sigma)\right)\prod_{i=1}^{n_1}(A_1)^{a_i}_{\alpha_{\tau(i)}}\times
$$
\bea
\times \prod_{j=1+n_1}^{n}(A_2)^{a_j}_{\alpha_{\tau(j)}}
\prod_{i=1}^{m_{1}}(B_1^\dagger)^{\alpha_i}_{a_{\sigma(i)}} 
\prod_{j=1+m_{1}}^{n}(B_2^\dagger)^{\alpha_j}_{a_{\sigma(j)}}
\label{firstslots}
\eea
$$
\equiv {\rm Tr} (P_{R,\{r\}}\, A_1^{\otimes n_{11}+n_{12}}A_2^{\otimes n_{21}+n_{22}}\,\tau\,
(B_1^\dagger)^{\otimes n_{11}+n_{21}}(B_2^\dagger)^{\otimes n_{12}+n_{22}})
$$ 
where
$$
P_{R,\{r\}} ={1\over n_{11}! n_{22}!n_{12}!n_{21}!}
\sum_{\sigma\in S_{n}}
{\rm Tr}_{\{r\}}\left(\Gamma_R (\sigma)\right)\sigma
$$
In the last line of (\ref{firstslots}) we have switched to a trace within $V^{\otimes n}$. The above explicit 
formula spells out how we are filling the ``slots'' from 1 to $n$ with the $A_i$s and $B^\dagger_j$s.
The specific way in which the slots are populated determines how the
subgroups $S_{n_1}\times S_{n_2}$ and $S_{m_1}\times S_{m_2}$ are embedded into $S_n$.
The operator $\tau$ dictates how the $A_i$s and $B^\dagger_j$s are to be combined to produce $\phi_{ij}$s. 
The specific $\tau$ we must choose to achieve a specific joining will not in general be unique. It would
also be possible to replace $\tau$ by some more general element of the group algebra\footnote{The $n_{ij}$ continue
to count the number of boxes in the Young diagrams $r_{ij}$, but no longer give the number 
of composite scalars $(\phi_{ij})^{a}_{b}$ from which the operator is built.}. We will pursue
this possibility below. The name ``restricted Schur polynomial,'' regardless of the specific $\tau$ used in the 
construction, reflects that fact that for all of these operators the index structure associated with the $U(N)$ group on 
which the projector $P_{R,\{r\}}$ acts is organized using the symmetric group and its subgroups. It is now a 
simple exercise to show that
\begin{center}
  \fbox{
    \begin{minipage}[c]{14cm}
      \small{
        {\vskip 0.05cm}
        \bea
           \langle
           O_{R,\{r\}}O^\dagger_{S,\{s\}} \rangle &&=\\
           \nonumber
           \sum_{\psi\circ\lambda\in \, S_{n_1}\times S_{n_2}}\,
           \sum_{\mu\circ\nu\in \, S_{m_1}\times S_{m_2}}&&
           \label{kkkeycorrelator}
           {\rm Tr}(P_{R,\{r\}} \, \psi\circ\lambda P_{S,\{s\}}\, \mu\circ\nu)\,\,
           {\rm Tr}(\tau^\dagger \, \psi^{-1}\circ\lambda^{-1} \, \tau \, \mu^{-1}\circ\nu^{-1} )
        \eea
        {\vskip 0.05cm}
      } 
    \end{minipage}
  }
\end{center}
The sum over $\psi\circ\lambda$ sums all possible Wick contractions between the $A_i$s and the sum over $\mu\circ\nu$ sums all possible Wick
contractions between the $B_i$s. After making a convenient choice for $\tau$ we will show how to evaluate (4.2) in general.

\subsection{Number of $A_i$s equal number of $B_i^\dagger$s; $n_{12}=n_{21}=0$}

In this subsection we consider the case that $n_1=m_1$, $n_2=m_2$ and further that
$n_{12}=0=n_{21}$. With this choice $\{ r\}= \{ r_{11},r_{22}\}$.
There are a number of nice simplifications that arise in this case. First, we may take 
$\tau$ to be the identity permutation. Secondly, both $P_{R,\{r\} }$ and $P_{S,\{ s\}}$
commute with all elements of $S_{n_{11}}\times S_{n_{22}}$. Thus, the two point correlator becomes
$$
\langle
O_{R,\{r\}}O^\dagger_{S,\{s\}}
\rangle
=\sum_{\psi\circ\lambda\in \, S_{n_{11}}\times S_{n_{22}}}\,
\sum_{\mu\circ\nu\in \, S_{n_{11}}\times S_{n_{22}}}
$$
$$
{\rm Tr}(P_{R,\{r\}} \, \psi\circ\lambda \, P_{S,\{s\}}\, \mu\circ\nu)
{\rm Tr}(\psi^{-1}\circ\lambda^{-1} \, \mu^{-1}\circ\nu^{-1} )
$$
$$
=\sum_{\psi\circ\lambda\in \, S_{n_{11}}\times S_{n_{22}}}\,
\sum_{\mu\circ\nu\in \, S_{n_{11}}\times S_{n_{22}}}
{\rm Tr}(P_{R,\{r\}} \, P_{S,\{s\}}\, \psi\mu\circ\lambda\nu)
{\rm Tr}( (\psi\mu)^{-1}\circ (\lambda\nu)^{-1})
$$
$$
=n_{11}!n_{22}!
\sum_{\psi\circ\lambda\in \, S_{n_{11}}\times S_{n_{22}}}\,
{\rm Tr}(P_{R,\{r\}} \, P_{S,\{s\}}\, \psi\circ\lambda)
{\rm Tr}( \psi^{-1}\circ \lambda^{-1})\, .
$$
Next we use the identity
$$
P_{R,\{r\}} P_{S,\{s\}} = \delta_{RS}\delta_{\{r\},\{s\}}
{n!\over n_{11}!n_{22}! d_R}P_{R,\{r\}}
$$
proved in \cite{Bhattacharyya:2008rb}, and the identity (in this next formula ${\rm Tr}_n$ 
denotes a trace over $V^{\otimes n}$, $f_s$ is a product of the
factors of Young diagram $s$ and ${\rm hooks}_s$ is a product of the hook lengths of Young diagram $s$)
$$
{\rm Tr}_{n}(\psi\circ \lambda)={\rm Tr}_{n_{11}}(\psi){\rm Tr}_{n_{22}}(\lambda)
=\sum_{s\vdash n_{11}}\chi_s(\psi){f_s\over {\rm hooks}_s} 
 \sum_{t\vdash n_{22}}\chi_t(\lambda){f_t\over {\rm hooks}_t}
$$
which follows as a consequence of Schur-Weyl duality, to obtain
$$
\langle
O_{R,\{r\}}O^\dagger_{S,\{s\}}
\rangle
={n!\over d_R}\delta_{RS}\delta_{\{r\},\{s\}}
\sum_{\psi\circ\lambda\in \, S_{n_{11}}\times S_{n_{22}}}\times$$
$$\times
\sum_{u\vdash n_{11}}\chi_u(\psi){f_u\over {\rm hooks}_u} 
\sum_{t\vdash n_{22}}\chi_t(\lambda){f_t\over {\rm hooks}_t}
{\rm Tr}(P_{R,\{r\}} \, \psi\circ\lambda)\, .
$$
To do this sum, note that
$$
\sum_{\psi\in S_{n_{11}}} \chi_s(\psi)\psi = {n_{11}!\over d_s}P_s\qquad
\sum_{\lambda\in S_{n_{22}}} \chi_t(\lambda)\lambda = {n_{22}!\over d_t}P_t
$$
where $P_s$ and $P_t$ are correctly normalized projectors, projecting to the irrep $s$ of $S_{n_{11}}$ and $t$ of
$S_{n_{22}}$ respectively. Thus,
$$
\langle
O_{R,\{ r\}}O^\dagger_{S,\{ s\}}
\rangle
=\sum_u\sum_t{n!n_{11}!n_{22}! f_u f_t\over d_R d_u d_t {\rm hooks}_u{\rm hooks}_t}\delta_{RS}\delta_{\{r\},\{s\}} 
{\rm Tr}(P_{R,\{r\}} \, P_u P_t)
$$
$$
={n!n_{11}!n_{22}! f_{r_{11}} f_{r_{22}}\over d_R d_{r_{11}} d_{r_{22}} {\rm hooks}_{r_{11}}{\rm hooks}_{r_{22}}}
\delta_{RS}\delta_{\{ r\},\{s\}} {\rm Tr}(P_{R,\{r\}})
$$
$$
=\delta_{RS}\delta_{\{r\},\{s\}}
{{\rm hooks}_R f_{r_{11}} f_{r_{22}}f_R\over{\rm hooks}_{r_{11}}{\rm hooks}_{r_{22}}}
$$
To obtain the final result we used the value of ${\rm Tr}(P_{R,\{ r\}})$ which has been computed in \cite{de Mello Koch:2007uu}.

The basic result of the subsection is
\begin{center}
  \fbox{
    \begin{minipage}[c]{14cm}
      {\vskip 0.05cm}
      \bea
         \langle
         O_{R,\{ r\}}O^\dagger_{S,\{ s\}}
         \rangle
         =\delta_{RS}\delta_{\{r\},\{s\}}
         {{\rm hooks}_R f_{r_{11}} f_{r_{22}}f_R\over{\rm hooks}_{r_{11}}{\rm hooks}_{r_{22}}}
         \label{firstcorr}
      \eea
      {\vskip 0.05cm}
    \end{minipage}
  }
\end{center}

\subsection{$n_2=0$, $n_1=m_1+m_2$}
\label{closedunderD}

With this choice $\{ r\}= \{ r_{11},r_{12}\}$.
There are again a number of nice simplifications that arise in this case. First, we may again take 
$\tau$ to be the identity permutation. Secondly, both $P_{R,\{r\} }$ and $P_{S,\{ s\}}$
commute with all elements of $S_{n_{11}}\times S_{n_{12}}$. Thus, the two point correlator becomes
$$
\langle
O_{R,\{r\}}O^\dagger_{S,\{s\}}
\rangle
=\sum_{\sigma\in \, S_{n}}\,
\sum_{\rho\in \, S_{m_1}\times S_{m_2}}
{\rm Tr}(P_{R,\{r\}} \, \sigma \, P_{S,\{s\}}\, \rho)
{\rm Tr}(\sigma^{-1}\rho^{-1} )
$$
$$
=m_1!m_2!\sum_{\sigma\in \, S_{n}}\,
{\rm Tr}(P_{R,\{r\}} \, P_{S,\{s\}}\, \sigma)
{\rm Tr}( \sigma^{-1})
$$
$$
={n!\over d_R}\delta_{RS}\delta_{\{r\},\{s\}}
\sum_{\sigma\in \, S_{n}}\,{\rm Tr}(P_{R,\{r\}}\, \sigma) {\rm Tr}( \sigma^{-1})
$$
$$
={n!\over d_R}\delta_{RS}\delta_{\{r\},\{s\}}
\sum_{T\vdash n}{f_T\over {\rm hooks}_T}
\sum_{\sigma\in \, S_{n}}\,{\rm Tr}(P_{R,\{r\}}\, \sigma) \chi_T ( \sigma^{-1})
$$
$$
={n!\over d_R}\delta_{RS}\delta_{\{r\},\{s\}}
\sum_{T\vdash n}{f_T\over {\rm hooks}_T}{n!\over d_T} {\rm Tr}(P_{R,\{r\}}\, P_T)
$$
$$
={f_R n!\over d_R}\delta_{RS}\delta_{\{r\},\{s\}} {\rm Tr}(P_{R,\{r\}})
$$
$$
={(f_R)^2 n!\over d_R{\rm hooks}_{r_{11}}{\rm hooks}_{r_{12}}}\delta_{RS}\delta_{\{r\},\{s\}}
$$

The basic result of the subsection is
\begin{center}
  \fbox{
    \begin{minipage}[c]{14cm}
      {\vskip 0.05cm}
      \bea
         \langle
         O_{R,\{ r\}}O^\dagger_{S,\{ s\}}
         \rangle
         =\delta_{RS}\delta_{\{r\},\{s\}}
         {{\rm hooks}_R (f_R)^2 \over {\rm hooks}_{r_{11}}{\rm hooks}_{r_{12}}}
         \label{secondcorr}
      \eea
      {\vskip 0.05cm}
    \end{minipage}
  }
\end{center}

\subsection{General Case}
\label{generalcase}

In this section we will consider general $n_{ij}$. We will find it useful to allow $\tau$ to be a general
element of the group algebra. We will find it convenient to distribute the Higgs fields in the slots as follows
$$
O_{R,\{r\}}={1\over n_{11}! n_{22}!n_{12}!n_{21}!}
\sum_{\sigma\in S_{n}}
{\rm Tr}_{\{r\}}\left(\Gamma_R (\sigma)\right)\prod_{i=1}^{n_1}(A_1)^{a_i}_{\alpha_{i}}\prod_{j=1+n_1}^{n}(A_2)^{a_j}_{\alpha_{j}}
(\tau)^{\alpha_1\cdots \alpha_n}_{\beta_1\cdots\beta_n}\times
$$
\bea
\times 
\prod_{i=1}^{n_{11}}(B_1^\dagger)^{\beta_i}_{a_{\sigma(i)}} 
\prod_{i=1+n_{11}}^{n_{1}}(B_2^\dagger)^{\beta_i}_{a_{\sigma(i)}}
\prod_{i=1+n_1}^{n_1+n_{21}}(B_1^\dagger)^{\beta_i}_{a_{\sigma(i)}} 
\prod_{i=1+n_1+n_{21}}^{n}(B_2^\dagger)^{\beta_i}_{a_{\sigma(i)}}
\label{identifyslots}
\eea
Compare to (\ref{firstslots}) and notice that this is not the same distribution of the Higgs fields.
We will summarize this as
$$
  {\cal O}_{R,\{r\}}={\rm Tr}(P_{R,\{r\}} A^{\otimes n} \,\tau\, B^{\dagger\otimes n})
$$
where for simplicity, our notation does not spell out which fields inhabit which slots.
Standard manipulations give (this assumes a Hermittian $\tau$ which is the case we consider below)
$$
  \langle {\cal O}_{R,\{r\}} {\cal O}_{S,\{s\}}^\dagger \rangle =
  \sum_{\rho\in S_{m_1}\times S_{m_2}}\sum_{\sigma\in S_{n_1}\times S_{n_2}}
         {\rm Tr}(P_{R,\{r\}}\sigma P_{S,\{s\}}\rho){\rm Tr}(\tau \rho^{-1} \tau \sigma^{-1})\, .
$$
To see how the subgroups are embedded in $S_n$, note that $S_{n_1}\times S_{n_2}$ acts on slots occupied by the $A$s and
$S_{m_1}\times S_{m_2}$ acts on slots occupied by $B^\dagger$s. The formula (\ref{identifyslots}) clearly states how the
slots are populated. Note that on the right hand side there are two traces and the sums to be performed have one
element of the symmetric group in one trace and the inverse of this in the second trace. The corresponding computation for 
gauge group $U(N)$ has one trace and both an element of the symmetric group and its inverse, in the same trace. This is a 
key observation that motivates what follows.

We could reduce the above result to the corresponding result obtained for a $U(N)$ gauge group if we choose $\tau$ so that
\bea
  {\rm Tr}(\tau \rho^{-1} \tau \sigma^{-1})=\delta (\rho^{-1} \sigma^{-1})\, .
  \label{ndelta}
\eea
Define ($R\vdash n$)
$$
  \Phi_R ={d_R\over n!f_R}\sum_{\sigma\in S_n}\chi_R(\sigma^{-1})\sigma\, .
$$
A rather straight forward computation now gives
$$
  {\rm Tr}(\Phi_R \psi)={d_R\over n!}\chi_R(\psi)
$$
where the trace is over $V^{\otimes n}$. Recalling that the delta function on the symmetric group is
$$
  \delta (\sigma ) ={1\over n!}\sum_{R\vdash n} d_R\chi_R(\sigma )
$$
we have
$$
  {\rm Tr}(\sum_{R\vdash n}\Phi_R \psi)=\delta (\psi)\, .
$$
This motivates the choice
\bea
  \tau =\sum_{R\vdash n} {d_R\over n!\sqrt{f_R}}\sum_{\sigma\in S_n}\chi_R(\sigma^{-1})\sigma\, .
  \label{choosetau}
\eea
With this choice (\ref{ndelta}) holds so that
$$
  \langle {\cal O}_{R,\{r\}}{\cal O}_{S,\{s\}}^\dagger \rangle =\sum_{\sigma \in S_{n_1}\times S_{n_2}\cap S_{m_1}\times S_{m_2}}
      {\rm Tr}(P_{R,\{r\}}\sigma P_{S,\{s\}}\sigma^{-1})\, .
$$
Notice that $S_{n_1}\times S_{n_2}\cap S_{m_1}\times S_{m_2}=S_{n_{11}}\times S_{n_{12}}\times S_{n_{21}}\times S_{n_{22}}$. Thus,
$\sigma$ commutes with the projectors in the last equation. After summing over $\sigma$ we have
$$
  \langle {\cal O}_{R,\{r\}}{\cal O}_{S,\{s\}}^\dagger \rangle =n_{11}!n_{12}!n_{21}!n_{22}! {\rm Tr}(P_{R,\{r\}} P_{S,\{s\}})\, .
$$
A straight forward application of the results of \cite{Bhattacharyya:2008rb,de Mello Koch:2007uu} now gives
\begin{center}
  \fbox{
    \begin{minipage}[c]{14cm}
      {\vskip 0.05cm}
      \bea
         \langle
         O_{R,\{ r\}}O^\dagger_{S,\{ s\}}
         \rangle
         =\delta_{RS}\delta_{\{r\},\{s\}}
         {{\rm hooks}_R f_R \over {\rm hooks}_{r_{11}}{\rm hooks}_{r_{12}}{\rm hooks}_{r_{21}}{\rm hooks}_{r_{22}}}\, .
         \label{gencorr}
      \eea
      {\vskip 0.05cm}
    \end{minipage}
  }
\end{center}

This clearly shows that our operators diagonalize the two point function in the subspace of operators with fixed $n_{ij}$.
However, even after fixing $n_i,m_i$, we can still change the $n_{ij}$, by changing the way we populate the slots with the
$B^\dagger$s which corresponds to changing the way that we embed $S_{m_1}\times S_{m_2}$ in $S_n$. Projectors corresponding
to different $n_{ij}$ will not in general be orthogonal. However, in this case (\ref{ndelta}) is never satisfied so that the
operators continue to be orthogonal.

\section{Action of the Dilatation Operator}

The two loop dilatation generator in the sector with $n_2=0$ is \cite{Kristjansen:2008ib}
$$
  D=-\left( {4\pi\over k}\right)^2 :{\rm Tr}
     \Big[ \left(B_2^\dagger A_1 B_1^\dagger - B_1^\dagger A_1 B_2^\dagger\right)\left(
            {\partial\over\partial B_2^\dagger}{\partial\over\partial A_1}{\partial\over\partial B_1^\dagger} 
         -  {\partial\over\partial B_1^\dagger}{\partial\over\partial A_1}{\partial\over\partial B_2^\dagger}\right)\Big]:\, .
$$ 
We will compute the action of the dilatation operator on operators normalized so that
$$
\langle \hat{O}_{R,\{ r\}} \hat{O}_{S,\{ s\}}^\dagger\rangle = f_R\delta_{RS}\delta_{\{ r\}\{ s\}}
\, .
$$
This choice of normalization makes the present problem look as similar as possible to that of \cite{VinceKate}.
The relation of these normalized operators (indicated with a hat) to the operators of section \ref{closedunderD} is
$$
O_{R,\{ r\}}=\sqrt{{\rm hooks}_R\, f_R\over {\rm hooks}_{r_{11}}\, {\rm hooks}_{r_{12}}}\hat{O}_{R,\{ r\}}\, .
$$
Using the methods of \cite{Koch:2010gp,VinceKate}, it is straight forward to obtain
$$
  D \hat{O}_{R,\{ r\}} =\sum_{S,\{s\}}M_{R,\{ r\}, S,\{ s\}}\hat{O}_{S,\{ s\}}
$$
where
{\small
$$
  M_{R,\{ r\}, S,\{ s\}} = 
\sqrt{{\rm hooks}_S\, f_S\, {\rm hooks}_{r_{11}}\, {\rm hooks}_{r_{12}}\over {\rm hooks}_R\, f_R\, {\rm hooks}_{s_{11}}\,{\rm hooks}_{s_{12}}}
\sum_{R'} {m_1\, m_2\, d_S\, c_{RR'} \over d_{s_{11}}\, d_{s_{12}}\, n\, d_{R'}}\times
$$
$$
\times\Big[
N {\rm Tr}\left( I_{S'R'}\big[ \Gamma_R \left((1,m_2+1)\right),P_{R,\{ r\}}\big]I_{R'S'} C\right)+
$$
$$
+{\rm Tr}\left( I_{S'R'}\big[ \Gamma_R \left((1,m_2+1)\right)P_{R,\{ r\}} \Gamma_R \left((1,m_2+1)\right)-P_{R,\{ r\}}\big]I_{R'S'} C\right)+
$$
$$
+(m_1-1){\rm Tr}\left( I_{S'R'}\big[ \Gamma_R \left((1,m_2+1)\right)P_{R,\{ r\}} \Gamma_R \left((1,m_2+2)\right)
                     -\Gamma_S \left((m_2+2,1,m_2+1)\right)P_{R,\{ r\}}\big]I_{R'S'} C\right)+
$$
$$
+(m_2-1){\rm Tr}\left( I_{S'R'}\big[ \Gamma_R \left((1,m_2+1)\right)P_{R,\{ r\}} \Gamma_R \left((1,2)\right)
              -P_{R,\{ r\}}\Gamma_S \left((2,1,m_2+1)\right)\big]I_{R'S'}
C \right)\Big]
$$
}
and
$$  C= \big[ P_{S,\{ s\}},\Gamma_S \left((1,m_2+1)\right)\big]\, . $$
To obtain this result we have used the first $m_2$ slots for the $\phi_{12}$s and the next $m_1$ slots for the $\phi_{11}$s.

We will study the spectrum of anomalous dimensions for operators whose labels $R,\{ r\}$ are all Young diagrams with two long rows.
In this case we can use $U(2)$ group theory to construct the projectors as explained in \cite{bhw,dgm}. The extension to operators 
whose labels $R,\{ r\}$ are all Young diagrams with $p$ long rows is also possible; in this case $U(p)$ group theory is used\cite{dgm}.
To set up the two long rows problem, we will employ a more convenient labeling for our operators. Notice that $m_1$ and $m_2$ are fixed.
We study the limit in which both $m_1$ and $m_2$ are $O(N)$, but $m_2\ll m_1$.
Denote the number of boxes in row 1 of $r_{12}$ minus the number of boxes in row 2 by $2j$. The number of boxes in row 2 is thus 
${m_2-2j\over 2}$. For $m_2=24$ and $j=4$ we have
$$
  r_{12}=\yng(16,8)
$$
In this way, we trade $r_{12}$ for an integer $j$. Next, imagine that to obtain $r_{11}$ from $R$ we need to pull $\nu_1$ boxes from the
first row of $R$ and $\nu_2$ boxes from the second row of $R$. Since we know that $\nu_1+\nu_2=m_2$ it is enough to specify 
$\nu_1-\nu_2\equiv 2j^3$. Finally, we will trade $r_{11}$ for the two integers $b_0$ and $b_1$. $b_1$ is the number of columns with a single
box while $b_0$ is the number of columns containing two boxes. Note that this notation is redundant because $2b_0+b_1=m_1$.
Thus, we trade the three Young diagrams $R,r_{11},r_{12}$ for the integers $b_0,b_1,j,j^3$. See Figure 1
for a summary. Using the ideas developed in \cite{bhw,dgm} we find after a straight forward but tedious computation
\bea
  D \hat{O}_{j,j^3} (b_0,b_1)
        && =\left({4\pi\over k}\right)^2\left[ \left(-{N\over 2}\left( m_2-{(m_2+2)(j^3)^2\over j(j+1)}\right)\right.\right.\\ 
&&\left. -{m_2^2\over 4}+m_2+j_3^2-j(j+1)-{j_3^2 (4-m_2^2)\over 4j(j+1)}\right)\Delta \hat{O}_{j,j^3}(b_0,b_1)\nonumber
\eea
$$
+ N\sqrt{(m_2+2j+4)(m_2-2j)\over (2j+1)(2j+3)}{(j+j^3+1)(j-j^3+1)\over 2(j+1)}
\left(1+ {m_2 - 2j - 4\over 2N}\right)\Delta \hat{O}_{j+1,j^3}(b_{0},b_{1})
$$
$$
+\sqrt{(m_2+2j+2)(m_2-2j+2)\over (2j+1)(2j-1)}{(j+j^3 )(j-j^3 ) \over 2 j}\left.
\left(1+ {m_2-2j-2\over 2N}\right)\Delta \hat{O}_{j-1,j^3}(b_{0},b_{1})\right]
$$
where
\begin{eqnarray}
\Delta \hat{O}_{j,j^3}(b_{0},b_{1}) &&=\sqrt{(N+b_0)(N+b_0+b_1)}(\hat{O}_{j,j^3}(b_0+1,b_1-2)+\hat{O}_{j,j^3}(b_0-1,b_1+2))\nonumber \\
 &&-(2N+2b_0+b_1)\hat{O}_{j,j^3}(b_0,b_1).\label{DeltaDefn}
\end{eqnarray}
This is remarkably similar to the result obtained for two row restricted Schurs in the $SU(2)$ sector of ${\cal N}=4$ super Yang Mills
theory\cite{bhw}. In particular, the fact that only the combination $\Delta O_{j,j^3}(b_{0},b_{1})$ appears implies that after we have 
diagonalized on the $j$ label, the problem of diagonalizing on the $b_0,b_1$ labels again reduces to diagonalizing a set of decoupled 
harmonic oscillators.

\begin{figure}[h]
   \centering
   \label{fig:newlab}
   {\epsfig{file=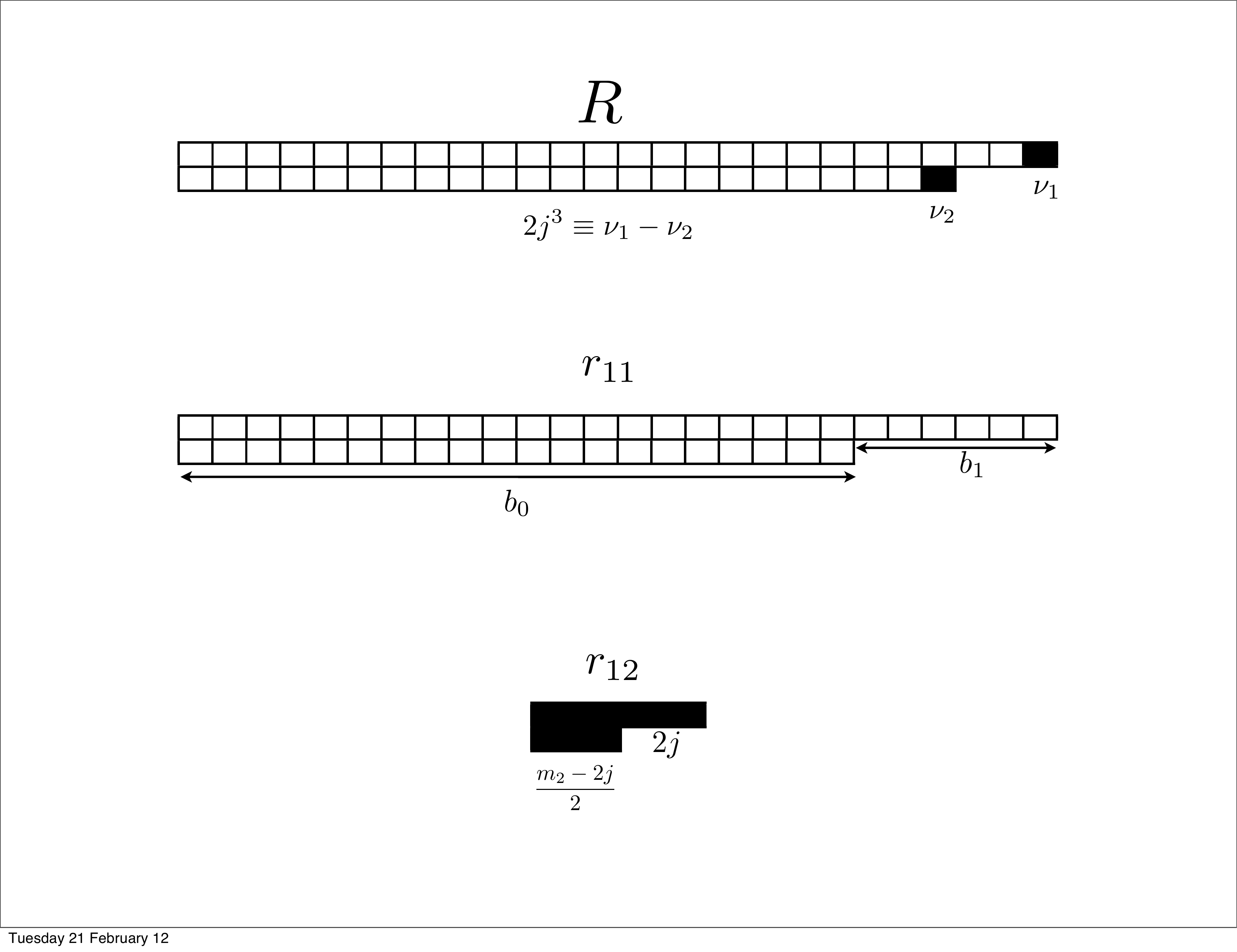,width=11.5cm,height=9.0cm}}
   \caption{A summary of the $U(2)$ labeling}
 \end{figure}

\section{Spectrum of Anomalous Dimensions}
\label{scaling}

For the $SU(2)$ sector of ${\cal N}=4$ super Yang-Mills theory \cite{deMelloKoch:2011ci} have proved that studying the dilatation
operator in a continuum limit reproduces the spectrum obtained by solving the original discrete anomalous dimension eigenvalue
problem. In this section we will consider a continuum limit of the diltation operator that reduces to the problem studied in
\cite{bhw}.

Consider first the problem of diagonalizing on the $b_0,b_1$ labels. We introduce $x = 2{b_1\over \sqrt{N+b_0}}$. For any finite
arbitrarily large $x$ we have $b_1\sim\sqrt{N}$. In this limit \cite{bhw} show that the operator $\Delta$ defined in 
(\ref{DeltaDefn}) reduces to the harmonic oscillator Hamiltonian which is easily diagonalized. Consider next the problem of
diagonalizing on the $j,j_3$ labels. To solve this problem we will consider the double scaling limit defined by taking
$m_2\to\infty$, $b_1\to\infty$ holding ${m\over b_1}\sim\gamma\ll 1$ fixed. In this case $\sqrt{2\over m}j$ becomes a continuous
variable. It is straight forward to see that in this continuum limit, the action found for the dilatation operator in the previous
section reduces to the continuum limit of the action of the dilatation operator studied in \cite{bhw}. From the results of that
work we know that if $m_2=2n$ we obtain a set of oscillators with frequency $\omega_i$ and degeneracy $d_i$ given by
$$\omega_i =8iN\left({4\pi\over k}\right)^2,\qquad d_i=2(n-i)+1,\qquad i=0,1,...,n\, .$$
and if $m_2=2n+1$ we obtain a set of oscillators with 
frequency $\omega_i$ and degeneracy $d_i$ given by
$$\omega_i =8iN\left({4\pi\over k}\right)^2,\qquad d_i=2(n-i+1),\qquad i=0,1,...,n\, .$$

{\vskip 1.0cm}

\noindent
{\it Acknowledgements:} We would like to thank 
Pawel Caputa for enjoyable, helpful discussions. 
RdMK and BAEM are supported by the South African Research Chairs
Initiative of the Department of Science and Technology and National Research Foundation.
JM is supported by the National Research Foundation under the Thuthuka and Incentive Funding for Rated Researchers Program.
AP is supported by an NRF Innovation Postdoctoral Fellowship.
Any opinion, findings and conclusions or recommendations expressed in this material
are those of the authors and therefore the NRF and DST do not accept any liability
with regard thereto.

\appendix

\section{More General Gauge Groups}

The operators we have constructed above also give a complete basis for gauge groups with an arbitrary number of factors
$U(N)\times U(N)\times \cdots \times U(N)$. We will illustrate the example of three factors $U(N)\times U(N)\times U(N)$.
Denote the indices associated with the gauge groups by $a,\alpha,A$. The theory is assumed to have three sets of fields,
all transforming in different bi fundamental representations of the factors
$$
  (A_i)^a_\alpha\, ,\qquad (B_i)^\alpha_A\, ,\qquad (C_i)^A_a\, .
$$
The most general operator in the theory can be written as a product of traces of the operators
$$
  (\phi_{ijk})^a_b\equiv (A_i B_j C_k )^a_b\, .
$$
We will assume that $i$ runs from 1 to $n_A$, $j$ from 1 to $n_B$ and $k$ from 1 to $n_C$. The number of
$\phi_{ijk}$ fields will be denoted by $n_{ijk}$. Repeating the arguments of section \ref{completeops}, 
it is clear that the restricted Schur polynomials are constructed by subducing 
$S_{n_{111}}\times S_{n_{112}}\times\cdots\times S_{n_A n_B n_C}$ irreps from $S_n$ irreps where 
$n=\sum_{ijk}n_{ijk}$. Thus, $\{r\}$ is a set of $n_A n_B n_C$ Young diagrams.
Our operators ($\tau$ is chosen as in (\ref{choosetau}))
$$
  O_{R,\{r\}} = {\rm Tr}(P_{R,\{ r\}}A^{\otimes n}\tau B^{\otimes n}\tau C^{\otimes n})
$$
have two point function
\bea
  \langle O_{R,\{ r\}}O^\dagger_{S,\{ s\}} \rangle =\delta_{RS}\delta_{\{r\},\{s\}}
  {{\rm hooks}_R f_R \over {\rm hooks}_{\{ r\}}}\, .
\eea
where ${\rm hooks}_{\{ r\}}$ is a product of hook factors, one for each of the $n_A n_B n_C$ Young diagrams
appearing in $\{ r\}$.


\begin{thebibliography}{30}
\parskip-2pt

\bibitem{Maldacena:1997re}
  J.~M.~Maldacena,
  ``The large N limit of superconformal field theories and supergravity,''
  Adv.\ Theor.\ Math.\ Phys.\  {\bf 2}, 231 (1998)
  [Int.\ J.\ Theor.\ Phys.\  {\bf 38}, 1113 (1999)]
  [arXiv:hep-th/9711200];\\
  S.~S.~Gubser, I.~R.~Klebanov and A.~M.~Polyakov,
  ``Gauge theory correlators from non-critical string theory,''
  Phys.\ Lett.\ B {\bf 428}, 105 (1998)
  [arXiv:hep-th/9802109];\\
  E.~Witten,
  ``Anti-de Sitter space and holography,''
  Adv.\ Theor.\ Math.\ Phys.\  {\bf 2}, 253 (1998)
  [arXiv:hep-th/9802150].

\bibitem{ABJM}
O.~Aharony, O.~Bergman, D.~L.~Jafferis and J.~Maldacena, 
``N=6 superconformal Chern-Simons-matter theories, M2-branes and their gravity duals,''
JHEP {\bf 0810}, 091 (2008)
[arXiv:0806.1218 [hep-th]].

\bibitem{BL}
J.~Bagger and N.~Lambert, 
``Modeling Multiple M2's,''
Phys.\ Rev.\ D {\bf 75}, 045020 (2007)
[hep-th/0611108],\\
A.~Gustavsson, 
``Algebraic structures on parallel M2-branes,''
Nucl.\ Phys.\ B {\bf 811}, 66 (2009)
[arXiv:0709.1260 [hep-th]],\\
J.~Bagger and N.~Lambert, 
``Gauge symmetry and supersymmetry of multiple M2-branes,''
Phys.\ Rev.\ D {\bf 77}, 065008 (2008)
[arXiv:0711.0955 [hep-th]],\\
J.~Bagger and N.~Lambert, 
``Comments on multiple M2-branes,''
JHEP {\bf 0802}, 105 (2008)
[arXiv:0712.3738 [hep-th]],\\
A.~Gustavsson, 
``One-loop corrections to Bagger-Lambert theory,''
Nucl.\ Phys.\ B {\bf 807}, 315 (2009)
[arXiv:0805.4443 [hep-th]].



\bibitem{Balasubramanian:2001nh}
  V.~Balasubramanian, M.~Berkooz, A.~Naqvi and M.~J.~Strassler,
  ``Giant gravitons in conformal field theory,''
  JHEP {\bf 0204}, 034 (2002)
  [arXiv:hep-th/0107119].

\bibitem{Corley:2001zk}
  S.~Corley, A.~Jevicki and S.~Ramgoolam,
  ``Exact correlators of giant gravitons from dual N = 4 SYM theory,''
  Adv.\ Theor.\ Math.\ Phys.\  {\bf 5}, 809 (2002)
  [arXiv:hep-th/0111222].

\bibitem{Balasubramanian:2004nb}
  V.~Balasubramanian, D.~Berenstein, B.~Feng and M.~x.~Huang,
  ``D-branes in Yang-Mills theory and emergent gauge symmetry,''
  JHEP {\bf 0503}, 006 (2005)
  [arXiv:hep-th/0411205].

\bibitem{de Mello Koch:2007uu} 
  R.~de Mello Koch, J.~Smolic and M.~Smolic,
  ``Giant Gravitons - with Strings Attached (I),''
  JHEP {\bf 0706}, 074 (2007)
  [hep-th/0701066].

\bibitem{Kimura:2007wy}
  Y.~Kimura and S.~Ramgoolam,
  ``Branes, Anti-Branes and Brauer Algebras in Gauge-Gravity duality,''
  arXiv:0709.2158 [hep-th].

\bibitem{Kimura:2009wy}
 Y.~Kimura,
  ``Non-holomorphic multi-matrix gauge invariant operators based on Brauer
  algebra,''
  arXiv:0910.2170 [hep-th].

\bibitem{Brown:2007xh}
  T.~W.~Brown, P.~J.~Heslop and S.~Ramgoolam,
  ``Diagonal multi-matrix correlators and BPS operators in N=4 SYM,''
  arXiv:0711.0176 [hep-th].

\bibitem{Brown:2008rr}
  T.~W.~Brown, P.~J.~Heslop and S.~Ramgoolam,
  ``Diagonal free field matrix correlators, global symmetries and giant
  gravitons,''
  arXiv:0806.1911 [hep-th].

\bibitem{jurgis}
  J.~Pasukonis and S.~Ramgoolam,
  ``From counting to construction of BPS states in N=4 SYM,''
  arXiv:1010.1683 [hep-th].

\bibitem{dey}
T.~K.~Dey, 
``Exact Large $R$-charge Correlators in ABJM Theory,''
JHEP {\bf 1108}, 066 (2011)
[arXiv:1105.0218 [hep-th]],\\
S.~Chakrabortty and T.~K.~Dey, 
``Correlators of Giant Gravitons from dual ABJ(M) Theory,''
arXiv:1112.6299 [hep-th].

\bibitem{Koch:2010gp}
  R.~d.~M.~Koch, G.~Mashile and N.~Park,
  ``Emergent Threebrane Lattices,''
  Phys.\ Rev.\  D {\bf 81}, 106009 (2010)
  [arXiv:1004.1108 [hep-th]].

\bibitem{VinceKate}
  V.~De~Comarmond, R.~de~Mello~Koch and K.~Jefferies,
  ``Surprisingly Simple Spectra,''
  [arXiv:1012.3884v1 [hep-th]].

\bibitem{bhw}
  W.~Carlson, R.~d.~M.~Koch and H.~Lin,
  ``Nonplanar Integrability,''
  arXiv:1101.5404 [hep-th].

\bibitem{Koch:2011jk}
  R.~d.~M.~Koch, B.~A.~E.~Mohammed, S.~Smith,
  ``Nonplanar Integrability: Beyond the SU(2) Sector,''
  [arXiv:1106.2483 [hep-th]].

\bibitem{dgm}
R.~d.~M.~Koch, M.~Dessein, D.~Giataganas, C.~Mathwin, 
``Giant Graviton Oscillators,'' [arXiv:1108.2761 [hep-th]]. 

\bibitem{deMelloKoch:2011ci} 
  R.~de Mello Koch, G.~Kemp and S.~Smith,
  ``From Large N Nonplanar Anomalous Dimensions to Open Spring Theory,''
  arXiv:1111.1058 [hep-th].

\bibitem{Minahan:2002ve}
  J.~A.~Minahan and K.~Zarembo,
  ``The Bethe-ansatz for N = 4 super Yang-Mills,''
  JHEP {\bf 0303}, 013 (2003)
  [arXiv:hep-th/0212208].

\bibitem{Beisert:2010jr}
  N.~Beisert {\it et al.},
  ``Review of AdS/CFT Integrability: An Overview,''
  arXiv:1012.3982 [hep-th]. For material which is very
  relevant, see in particular:\\
  C.~Kristjansen, 
  ``Review of AdS/CFT Integrability, Chapter IV.1: Aspects of Non-Planarity,'' 
  arXiv:1012.3997 [hep-th],\\ 
  K.~Zoubos,
  ``Review of AdS/CFT Integrability, Chapter IV.2: Deformations, Orbifolds and Open Boundaries,''
  arXiv:1012.3998 [hep-th].

\bibitem{McGreevy:2000cw}
  J.~McGreevy, L.~Susskind and N.~Toumbas,
  ``Invasion of the giant gravitons from anti-de Sitter space,''
  JHEP {\bf 0006}, 008 (2000)
  [arXiv:hep-th/0003075];\\
  M.~T.~Grisaru, R.~C.~Myers and O.~Tafjord,
  ``SUSY and Goliath,''
  JHEP {\bf 0008}, 040 (2000)
  [arXiv:hep-th/0008015];\\
  A.~Hashimoto, S.~Hirano and N.~Itzhaki,
  ``Large branes in AdS and their field theory dual,''
  JHEP {\bf 0008}, 051 (2000)
  [arXiv:hep-th/0008016].

\bibitem{Jeff}
D.~Giovannoni, J.~Murugan and A.~Prinsloo, 
``The giant graviton on $AdS_{4} \times \mathbb{CP}^{3}$ - another step towards the emergence of geometry,''
JHEP {\bf 1112}, 003 (2011)
[arXiv:1108.3084 [hep-th]].

\bibitem{AdS4likegiants}
A.~Hamilton, J.~Murugan and A.~Prinsloo, 
``Lessons from giant gravitons on $AdS_{5}\times T^{1,1}$,''
JHEP {\bf 1006}, 017 (2010)
[arXiv:1001.2306 [hep-th]],\\
N.~Gutierrez, Y.~Lozano and D.~Rodriguez-Gomez, 
``Charged particle-like branes in ABJM,''
JHEP {\bf 1009}, 101 (2010)
[arXiv:1004.2826 [hep-th]],\\
Y.~Lozano, M.~Picos, K.~Sfetsos and K.~Siampos, 
``ABJM Baryon Stability and Myers effect,''
JHEP {\bf 1107}, 032 (2011)
[arXiv:1105.0939 [hep-th]].

\bibitem{jurgissanjaye}
  J.~Pasukonis and S.~Ramgoolam,
  ``Quantum States to Brane Geometries via Fuzzy Moduli Spaces of Giant Gravitons,''
  [arXiv:1201.5588 [hep-th]].

\bibitem{D2-dipoles}
T. Nishioka and T. Takayanagi, 
``Fuzzy Ring from M2-brane Giant Torus", 
JHEP. {\bf 0810}, 082 (2008), 
[arXiv:0808.2691 [hep-th]],\\
A. Hamilton, J. Murugan, A. Prinsloo and M. Strydom, 
``A note on dual giant gravitons in AdS$_{4}\times\mathbb{CP}^{3}$",
JHEP. {\bf 0409}, 132 (2009), 
[arXiv:0901.0009 [hep-th]]

\bibitem{Collins:2008gc} 
  S.~Collins,
  ``Restricted Schur Polynomials and Finite N Counting,''
  Phys.\ Rev.\ D {\bf 79}, 026002 (2009)
  [arXiv:0810.4217 [hep-th]].

\bibitem{Bhattacharyya:2008xy} 
R.~Bhattacharyya, R.~de Mello Koch and M.~Stephanou, 
``Exact Multi-Restricted Schur Polynomial Correlators,''
JHEP {\bf 0806}, 101 (2008)
[arXiv:0805.3025 [hep-th]].

\bibitem{Bhattacharyya:2008rb} 
  R.~Bhattacharyya, S.~Collins and R.~d.~M.~Koch,
  ``Exact Multi-Matrix Correlators,''
  JHEP {\bf 0803}, 044 (2008)
  [arXiv:0801.2061 [hep-th]].

\bibitem{Kristjansen:2008ib} 
  C.~Kristjansen, M.~Orselli and K.~Zoubos,
  ``Non-planar ABJM Theory and Integrability,''
  JHEP {\bf 0903}, 037 (2009)
  [arXiv:0811.2150 [hep-th]]. For ABJ theory see:\\
P.~Caputa, C.~Kristjansen and K.~Zoubos, 
``Non-planar ABJ Theory and Parity,''
Phys.\ Lett.\ B {\bf 677}, 197 (2009)
[arXiv:0903.3354 [hep-th]].

\end{thebibliography}
\end{document}